\documentclass[
superscriptaddress,
twocolumn,
amsmath,amssymb,
aps,
prl,
raggedbottom
]{revtex4-2}

\usepackage{graphicx,framed} 
\usepackage{dcolumn} 
\usepackage{bm} 
\usepackage{braket}
\usepackage{dsfont}
\usepackage{color}
\usepackage{amsthm}
\usepackage{hyperref} 
\usepackage[normalem]{ulem} 
\usepackage{qcircuit}       
\usepackage{amsthm}

\usepackage{comment}
\usepackage{bbm}
\usepackage{algorithm}
\usepackage[noend]{algpseudocode}
\algdef{SE}[DOWHILE]{Do}{doWhile}{\algorithmicdo}[1]{\algorithmicwhile\ #1}
\makeatletter
\def\algbackskip{\hskip-\ALG@thistlm}
\makeatother

\newcounter{superequation} 
\makeatletter
\@addtoreset{equation}{superequation}
\makeatother

\def\bea{\begin{eqnarray}} \def\eea{\end{eqnarray}}
\definecolor{lightblue}{RGB}{73,151,208}
\definecolor{crimson}{RGB}{140,41,53}

\hypersetup{
    colorlinks,
    linkcolor={crimson},
    citecolor={lightblue},
    urlcolor={lightblue}
}

\theoremstyle{definition}

\newcommand{\tr}{\mathrm{Tr}}
\newcommand{\opket}[1]{|#1)}
\newcommand{\opbra}[1]{(#1|}
\newcommand{\opbraket}[2]{(#1|#2)}
\newcommand{\mL}{\mathcal{L}}

\newcommand{\mO}{\mathcal{O}}

\newcommand{\tscramb}{t_{\ast}}

\newcommand{\KrylovSlope}{\theta_{K}}
\newcommand{\LanczosSlope}{\alpha}

\newcommand{\lsec}[1]{\textit{#1.---}}

\begin{document}


\def\titlename{Krylov Winding and Emergent Coherence in Operator Growth Dynamics}
\title{\titlename}

\author{Rishik Perugu}
\affiliation{Department of Physics and Astronomy, University of California, Irvine, Irvine, CA 92697, USA}
\author{Bryce Kobrin}
\affiliation{Google Quantum AI, Venice, CA, USA}
\author{Michael O. Flynn}
\affiliation{BlocQ, Inc., 214 Homer Avenue, Palo Alto, CA 94301, USA}
\author{Thomas Scaffidi}
\email{tscaffid@uci.edu}
\affiliation{Department of Physics and Astronomy, University of California, Irvine, Irvine, CA 92697, USA}
\email{Emails}

\begin{abstract}
The operator wavefunction provides a fine-grained description of quantum chaos and of the irreversible growth of simple operators into increasingly complex ones.
Remarkably, at finite temperature this wavefunction can acquire a phase that increases linearly with the operator's size, a phenomenon called \emph{size winding}. Although size winding occurs naturally in a holographic setting, the emergence of a coherent phase in a scrambled operator remains mysterious from the standpoint of a thermalizing quantum many-body system. In this work, we elucidate this phenomenon by introducing the related concept of \textit{Krylov winding}, whereby the operator wavefunction acquires a phase which winds linearly with the Krylov index. We show that Krylov winding is a generic feature of quantum chaotic systems and is a direct consequence of the universal operator growth bound hypothesis. It gives rise to size winding under two additional conditions: (i) a low-rank mapping between the Krylov and size bases, which ensures phase alignment among operators of the same size, and (ii) the saturation of the ``chaos-operator growth'' bound $\lambda_L \leq 2 \alpha$ (with $\lambda_L$ the Lyapunov exponent and $\alpha$ the growth rate), which ensures a linear phase dependence on size. For systems which do not saturate this bound, with $h = \lambda_L / 2\alpha <1$, the winding with Pauli size $\ell$ becomes \emph{superlinear}, behaving as $\ell^{1/h}$. We illustrate these results with two classes of microscopic models: the Sachdev-Ye-Kitaev (SYK) model and its variants, and a disordered $k$-local spin model.
\end{abstract}

\maketitle

Recently, the emphasis in many-body dynamics has shifted from studying few-body correlation functions to investigating the fine-grained features of operator growth dynamics.
This shift has revealed fundamental insights into quantum chaos and information scrambling \cite{Hayden_2007, Sekino_2008,Maldacena_2016, Shenker_2014, shenker2015stringyeffectsscrambling,Hosur_2016, 
Roberts_2015, Keles2019, Zhou2020, Kobrin2021_SYKChaos}, emergent hydrodynamic behavior \cite{nahum2018operator,hyp,von2018operator, SwingleNature2018,Vedika, XuSwingleScrambling2024, Kim2021}, and quantum complexity~\cite{Roberts_2017,von2022operator, mi2021information,abanin2025constructive}. 
At the same time, the emergence of highly coherent quantum simulators provides unprecedented access to observables capturing the detailed structure of operator growth and scrambling \cite{garttner2017measuring,li2017measuring,mi2021information,abanin2025constructive}.

A central object in this pursuit is the ``operator wavefunction''.
At infinite temperature, the operator wavefunction is defined as $|O(t)) = \sum_P c_P(t) |P)$, where $O(t)$ is a time-evolved operator and $|P)$ is a complete operator basis, often the basis of Pauli strings.
The spreading of the wavefunction amplitude $|c_P(t)|$ encodes the growth of $O(t)$ in the space of Pauli strings---the hallmark of quantum scrambling.

Extending this notion to finite temperature is subtle, yet essential for understanding general properties of operator growth, including the universal bound on chaos~\cite{Maldacena_2016}.
A particularly natural choice is the ``asymmetrically thermal'' operator wavefunction~\cite{QuantumEpidemiology},
\bea
|\rho^{1/2} O(t))=\sum_{P} |c_{P}(t)|\,e^{i\phi_{P}(t)}\,|P),
\eea
with $\rho$ the thermal density matrix.
This definition is directly connected to physical observables, including the finite-temperature two-point function
$C(t)=\mathrm{Tr}[\rho\, O(t)O]=(\rho^{1/2}O\,|\,\rho^{1/2}O(t))$
which underlies linear response~\footnote{Here the inner product is $(A|B)=\mathrm{Tr}[A^\dagger B]$.}.
At finite temperature, the chaos bound $\lambda_L \le 2\pi k_B T$~\cite{Maldacena_2016} constrains the Lyapunov exponent $\lambda_L$, which characterizes the exponential rate at which the wavefunction amplitude $|c_{P}(t)|$ spreads toward operators of larger size.
Crucially, this wavefunction carries not only a magnitude but also a nontrivial complex \emph{phase}, $\phi_{P}(t)$.
This is unavoidable at finite temperature: even for a Hermitian observable $O$, the operator $\rho^{1/2}O(t)$ is non-Hermitian, necessitating complex coefficients in the Pauli basis.
Because $C(t)$ involves the overlap of the operator wavefunction with its initial state, its decay is governed not only by the spreading of the magnitude $|c_P|$ but crucially also by the interference of the phases $\phi_P$. 
The complex phase $\phi_P$ is thus a consequence of the fundamental non-Hermiticity of finite-temperature operator dynamics with a direct consequence on physical properties like the two-point function, and our goal is to understand its behavior under generic many-body quantum dynamics.

\begin{figure}[t!]
	\centering	
  \includegraphics[width=0.97\columnwidth]
 {{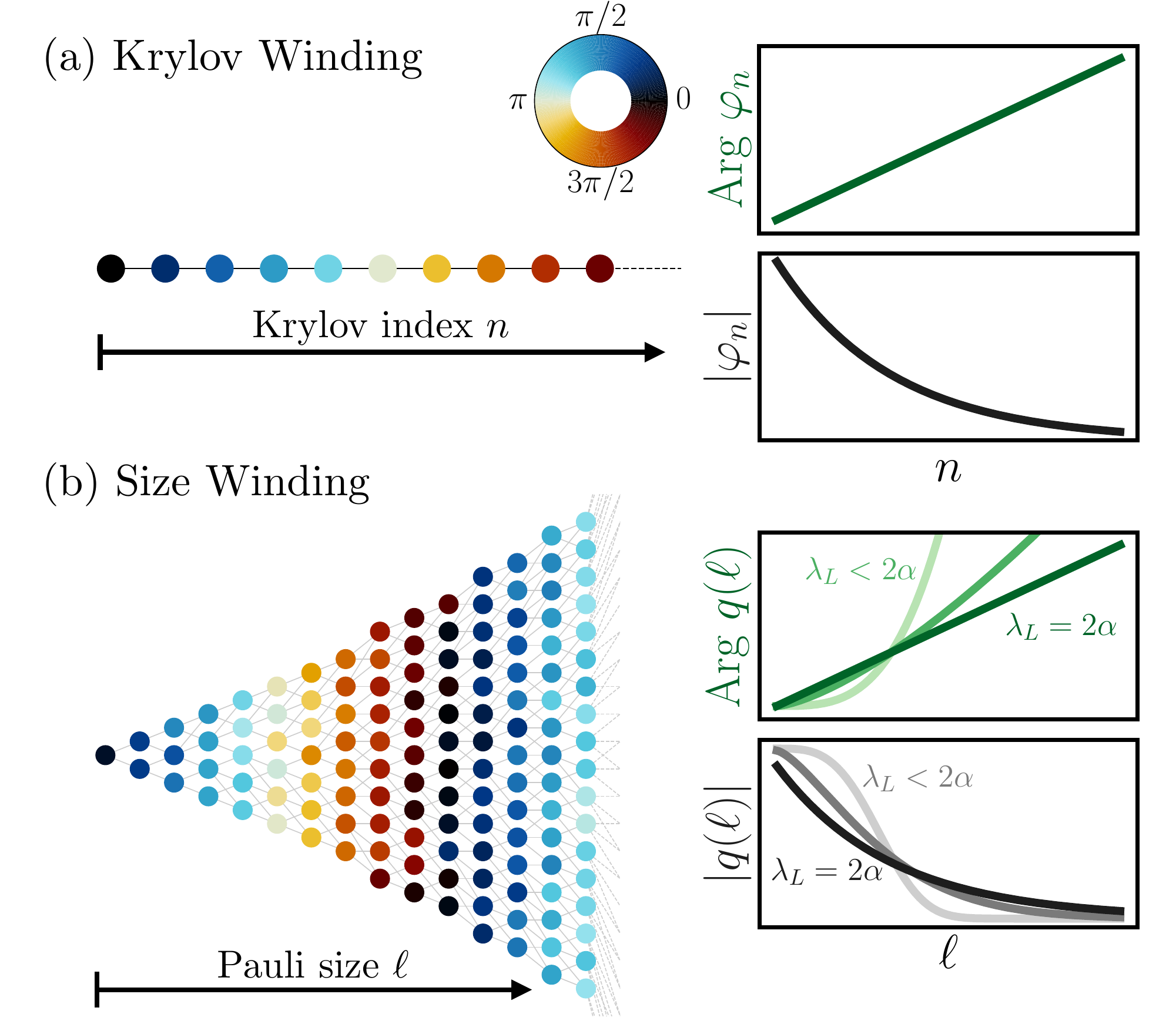}}
	\caption{ (a) Left: Sketch of the operator wavefunction in the one-dimensional Krylov basis. Colors give the complex phase of the Krylov wavefunction $\mathrm{Arg}[\varphi_n(t)] \sim n$ for a fixed time $t$. Right: Phase and magnitude of $\varphi_n$. The phase grows linearly and the magnitude decays exponentially with $n$, consistently with Eq.~\eqref{eq:varphilatetimes}. (b) Left: Sketch of the operator wavefunction in the Pauli basis. Each dot corresponds to a Pauli string, arranged by increasing size $\ell$ along the horizontal direction. The color on each dot shows the phase of the wavefunction $\mathrm{Arg}[c_P]$. Phase alignment is present but is in general only approximate: the phase is approximately constant in each size sector of fixed size $\ell$. Right: Sketch of the size winding distribution $q(l) = \sum_{P:|P|=l} c_P^2$ for different degrees of saturation of the chaos-operator growth bound $h = \lambda_L / 2 \alpha \leq 1$. The phase scales as $\ell^{1/h}$ and the magnitude decays as a compressed exponential $\sim \exp(-\ell^{1/h})$ (see Eq.~\eqref{Eq:qoflmain}).
}
\label{fig:Schematic}
\end{figure}

Remarkably, recent studies motivated by quantum gravity have revealed coherent behavior in this complex phase. 
Specifically, certain systems have been shown to exhibit a phenomenon, dubbed \emph{size winding}, in which $\phi_P$ increases linearly with the size of the operator $|P|$ \cite{brown2023quantum,nezami2023quantum},
\bea\label{eq:SizeWindingDef}
\phi_P(t) = \phi_0 + \theta(t) |P|
\eea
where, in the Pauli basis, $|P|$ is the number of non-identity operators in a Pauli string. 
Here $\theta(t)$ is the winding slope, which typically decays exponentially in time, and both $\theta(t)$ and $\phi_0$ are temperature dependent, as we will discuss further below.

From a holographic perspective, size winding provides an essential link between microscopic quantum dynamics and emergent geometry.
Within this framework, the winding slope, $\theta(t)$, is dual to the position of a particle propagating through a ``bulk'' spacetime~\cite{lin2019symmetries,brown2023quantum}.
Moreover, this duality is central to the recent discovery of traversable wormholes, which corresponds to a form of many-body quantum teleportation~\cite{gao2017traversable,maldacena2017diving,yoshida2017efficient,landsman2019verified,blok2021quantum,brown2023quantum,nezami2023quantum,schuster2022many}. 
While transmitting a scrambled state normally requires a finely tuned operation, it becomes simple in the presence of size winding, where it amounts to reversing the slope, $\theta(t) \rightarrow -\theta(t)$ \cite{brown2023quantum,nezami2023quantum,schuster2022many}. 
This teleportation mechanism has been proposed as a key signature for experimental investigations of quantum gravity \cite{gao2017traversable,maldacena2017diving,brown2023quantum,nezami2023quantum,schuster2022many,haehl2021six}.

Nevertheless, the emergence of size winding remains highly surprising from the perspective of thermalizing many-body systems: it requires that all Pauli operators of a given size share a common complex phase, and that this phase obeys a simple linear dependence on the size. 
Such coherence appears, at first sight, to be incompatible with the inherently irreversible nature of scrambling, in which simple operators evolve into increasingly complex ones over time.
Intriguingly, recent work has demonstrated size winding in a wide variety of quantum systems, from analytic calculations of all-to-all interacting systems \cite{ZhouSizeWinding,gao2024commuting} to numerical and experimental studies of small-sized, fully commuting models \cite{jafferis2022traversable,kobrin2023comment}. 
This evidence suggests that emergent coherence is a more widespread property of operator growth dynamics than initially suspected.

In this Letter, we propose a microscopic mechanism for such coherence by mapping the operator wavefunction to the Krylov basis, an ordered basis generated by the Liouvillian (Fig.~\ref{fig:Schematic}). 
In this framework, the dynamics reduces to a one-dimensional Schr\"odinger equation on a semi-infinite chain, where the operator wavefunction travels exponentially fast toward larger Krylov index $n$.
We introduce \emph{Krylov winding}---a linear dependence of the wavefunction phase on $n$---and demonstrate that it is a generic feature of interacting quantum systems arising from the linear growth of Lanczos coefficients~\cite{hyp}. 
Finally, we establish two sufficient conditions under which this Krylov coherence generates size winding. We illustrate these results in two microscopic models: a generalization of the Sachdev-Ye-Kitaev model~\cite{SY1992, SachdevSYK2015, kitaevtalk2015, SYK2016, KitaevSoftMode, ChowdhurySYKReview} introduced in Refs.~\cite{ ChenSYKplusSYKBath,Dissectingquantummanybodychaos}, and an all-to-all interacting spin system which we study numerically.

\begin{figure}[t!]
	\centering	
 \includegraphics[width=0.8\columnwidth]
  {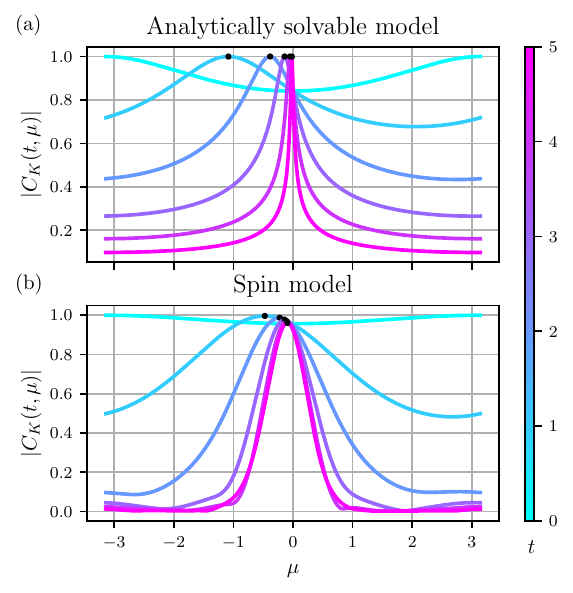}
   \includegraphics[width=0.8\columnwidth]
  {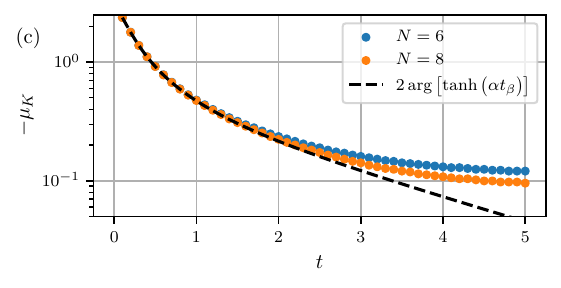}  
	\caption{(a),(b) Fourier transform $C_K(\mu,t)$ of the (squared) Krylov wavefunction for (a) the analytically solvable model of Eq.~\eqref{eq:ExactLinearWavefunction} with $\LanczosSlope=\pi\nu/\beta,\nu=0.5, \Delta=1/4$, and with time in units of $1/2\alpha$, and for (b) the spin model of Eq.~\eqref{eq:NonlocalHamiltonian} with $N=8$, averaged over 100 disorder realizations at $\beta=1$. In both cases, the distribution sharpens around a peak momentum $\mu_K$ (black dots) as time evolves. Curves are normalized such that $\sum_n |\varphi_n|^2 = 1$. (c) Peak momentum $\mu_K$ of the Krylov winding distribution $C_K(\mu,t)$ shown in Fig.~\ref{fig:Krylov_FTs}(b) versus time for the spin model of Eq.~\eqref{eq:NonlocalHamiltonian} at $\beta=1$. The evolution of $\mu_K$ at early times agrees with the analytical prediction before saturating at a time of order $\alpha^{-1} \log(N)$ due to finite size effects (see End Matter for more details on the saturation).}
\label{fig:Krylov_FTs}
\end{figure}

Another motivation for our work is to extend the Krylov formalism in a new direction. 
While the Krylov approach maps many-body dynamics to a one-dimensional Schrödinger equation, this equation has so far been restricted to real-valued wavefunctions, thereby prohibiting quantum interference effects. 
We demonstrate that finite-temperature dynamics inherently requires a \emph{complex} Krylov wavefunction, unlocking physics governed by such interference. 
This perspective reveals that the operator evolves into a right-moving wavepacket with an increasingly well-defined ``Krylov momentum,'' $\theta_K(t)$ (See Fig.~\ref{fig:Krylov_FTs}). 
This suggests an operational way of reversing the scrambling dynamics: applying the superoperator $e^{- 2 i \theta_K(t) n}$ could act as a ``momentum kick'' that reverses the wavepacket's direction, enabling an echo protocol directly in Krylov space which would generalize size-winding based teleportation.
Beyond its role in teleportation, phase winding offers a new diagnostic of scrambling itself. We will demonstrate that the phase scaling with size distinguishes maximal from sub-maximal chaos: saturation of the chaos-operator growth (COG) bound yields linear size winding, whereas sub-maximal chaos leads to a superlinear scaling [Fig.~\ref{fig:Schematic}(b)].

\lsec{Krylov Winding}
To define Krylov winding, we employ the Lanczos formalism for operator dynamics~\cite{ViswanathRecursion,hyp,xu2020does,PhysRevB.102.035147,Cao_2021,qi2023deephilbert,PhysRevB.110.155135,xiangyusaddledom,Nandy2025QuantumDynamics,Rabinovici2025KrylovComplexity, baiguera2025quantumcomplexitygravityquantum, XiangyuSizeConcentration, quantumscar, topologicaltransition, topologicalphases, integrabilitytochaos, K-ComplexityMBL, K-ComplexityGeometry, K-complexityCFT, SYKModel, KcomplexityandOperatorEntropy, K-complexitySU(2)Yang-Mills, afterScramblingTime,openquantumsystem, Krylovsuppressionofcomplexity, openquantumsystem2, generalizedKcomplexityToQuantumStates, KcomplexityOrthogonalPolynomials}. In the Hilbert space of operators, equipped with the infinite-temperature inner product $\opbraket{A}{B} = \tr(A^{\dagger}B)/\tr(\mathbbm{1})$, an operator $\opket{\mO}$ evolves under the Liouvillian superoperator $\mL=\left[H,\cdot\right]$ as $\opket{\mO(t)} = e^{i \mL t} \opket{\mO}$.

The Lanczos algorithm generates an orthonormal basis, the Krylov basis $\{\opket{O_n}\}$, in which the Liouvillian is tridiagonal. Starting from a normalized seed operator $\opket{O_{0}}$, the basis is constructed recursively:
\begin{equation}\label{eq:LanczosAlg}
\begin{aligned}
\opket{A_{n}} &= \mL\opket{O_{n-1}}-b_{n-1}\opket{O_{n-2}},\\
b_{n}&=\opbraket{A_{n}}{A_{n}}^{1/2},\quad
\opket{O_{n}}=\opket{A_{n}}/b_{n},
\end{aligned}
\end{equation}
with $O_{-1}=b_0=0$. The real numbers $b_n$ are the Lanczos coefficients. In this basis, $\mL\opket{O_{n}} = b_{n+1}\opket{O_{n+1}}+b_{n}\opket{O_{n-1}}$. By defining Hermitian basis operators $\opket{\mO_{n}} \equiv i^n \opket{O_{n}}$, the operator Krylov wavefunction $\varphi_{n}(t)\equiv \opbraket{\mO_{n}}{\mO(t)}$ obeys a simple nearest-neighbor hopping equation on a semi-infinite chain: 
\bea \label{hoppingmodel}
\partial_t\varphi_n = b_n\varphi_{n-1}-b_{n+1}\varphi_{n+1}.
\eea

As explained in the introduction we are interested in the operator $\rho_\beta^{1/2} \mO(t)$, with $\rho_{\beta}=e^{-\beta H}/\tr \left(e^{-\beta H}\right)$. In order to work with a Krylov basis of hermitian operators, we first use the following relation:
\begin{equation} \label{eq:complextimeevol}
\opket{\rho_{\beta}^{1/2}\mO(t)} = e^{i\mL(t+i\beta/4)} \opket{\rho_{\beta}^{1/4}\mO\rho_{\beta}^{1/4}}.
\end{equation}
We will thus seed the Lanczos algorithm with the symmetrized thermal operator $\opket{O_0} \propto \opket{\rho_{\beta}^{1/4}\mO\rho_{\beta}^{1/4}}$, and we will consider time evolution for a complex time $t_\beta \equiv t+i\beta/4$. (We note that the Krylov basis thus generated is the same as that used for finite-temperature Wightman 2-point functions~\cite{QuantumEpidemiology}).
The operator wavefunction in this Krylov basis is then given by
\begin{equation}\label{eq:KrylovExpansion}
\opket{\rho_{\beta}^{1/2}\mO(t)}=\sum_{n} \varphi_n(t_\beta)  \opket{\mO_n},
\end{equation}
where the coefficients $\varphi_n(t_\beta)$ are obtained by replacing $t \to t_\beta \equiv t+i\beta/4$ in the solution for $\varphi_n(t)$ from Eq.~\eqref{hoppingmodel}.

In analogy with size winding in Eq.~\eqref{eq:SizeWindingDef}, we define \textit{Krylov winding} as a linear relation between the phase of the Krylov wavefunction and the Krylov index $n$:
\begin{equation}
    \varphi_{n}(t_\beta) = |\varphi_{n}(t_\beta) | e^{i\KrylovSlope(t)n + i\theta_0(t)}, \quad \KrylovSlope(t),\theta_0(t)\in\mathbb{R}.
\end{equation}

\begin{figure}[t]
	\centering	
  \includegraphics[width=0.8\columnwidth]
 {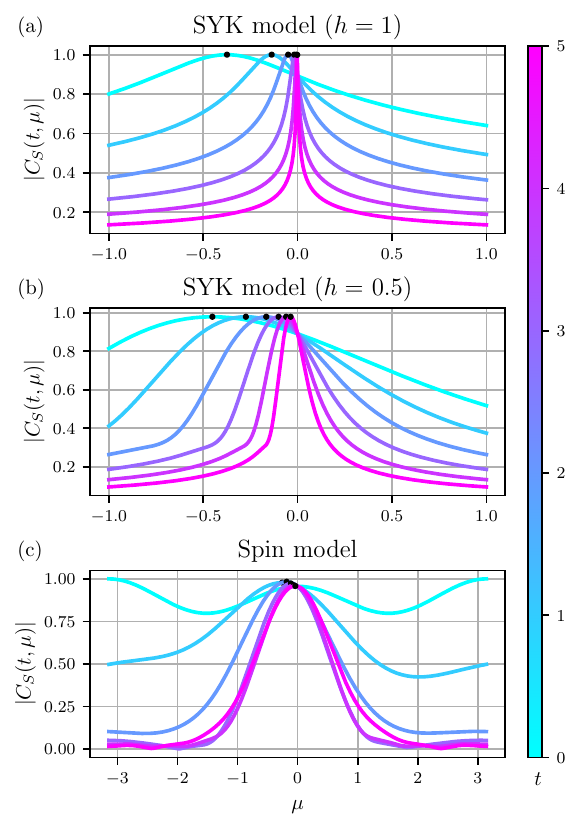}
	\caption{Fourier transform $C_S(\mu,t)$ of the size winding distribution for (a,b) large-$q$ SYK+bath model at $\nu = 0.5, q = 6, N = 3000$ with (a) $h=1$, (b) $h=0.5$, and (c) for the spin model of Eq.~\eqref{eq:NonlocalHamiltonian} with $N=8$, averaged over 100 disorder realizations at $\beta=1$. In all cases, the distribution develops a peak around a value $\mu_K(t)$ (black dots) which becomes sharper as time progresses.}
\label{fig:size_FTs}
\end{figure}

Krylov winding is a generic feature of non-integrable quantum systems, as we now explain. The operator growth hypothesis posits that for generic operators in such systems, the Lanczos coefficients grow linearly, $b_n \sim \alpha n$ for large $n$~\cite{hyp}. It is instructive to first consider the exactly solvable case $b_{n}=\LanczosSlope \sqrt{n(n+2\Delta-1)}$, for which the Krylov wavefunction is known analytically~\cite{hyp}. Applying the mapping $t\to t_\beta$ yields
\begin{equation}\label{eq:ExactLinearWavefunction}
\varphi_{n}(t_\beta)=\sqrt{\frac{\mathcal{N}\ \Gamma(2\Delta + n)}{\Gamma(n+1) \Gamma(2 \Delta)}}\frac{\tanh\left[\LanczosSlope t_\beta\right]^{n}}{\cosh\left[\LanczosSlope t_\beta\right]^{2\Delta}},
\end{equation}
where $\mathcal{N} = (\rho_{\beta}^{1/4}\mO\rho_{\beta}^{1/4} | \rho_{\beta}^{1/4}\mO\rho_{\beta}^{1/4})$ is a normalization factor. The phase of $\varphi_{n}$ winds linearly with $n$ with a slope
\begin{equation}\label{eq:alphaK}
\KrylovSlope(t) = \mathrm{Arg}[\tanh(\LanczosSlope t_\beta)] = \tan^{-1}\left(\frac{\sin(\LanczosSlope\beta/2)}{\sinh(2\LanczosSlope t)}\right).
\end{equation}
This Krylov winding slope depends on temperature both explicitly through the $\beta$ factor, but also through the temperature dependence of $\alpha$, which is model-dependent.
However, in the limiting case of models saturating the universal bound $\alpha \leq \pi /\beta$~\cite{hyp} (e.g. low-temperature SYK), the slope simplifies to $\theta_K(t) = \tan^{-1}(1/\sinh(2 \pi \beta^{-1} t))$.

More generally, for any system with $b_n \sim \alpha n$, the Krylov wavefunction for large enough $n$ behaves as a decaying exponential in $n$ with a width that grows exponentially in time: $\varphi_n(t) \sim \exp(-2n e^{-2\alpha t})$~\cite{hyp}. 
The replacement $t \to t_\beta$ thus gives 
\bea\label{eq:varphilatetimes}
\varphi_n(t_\beta) \sim \exp\left(-2n e^{-2\alpha t} \cos(\alpha \beta / 2)\right) \exp\left(i \theta_K(t) n \right)
\eea
with $\KrylovSlope(t) = 2e^{-2 \alpha t}\sin(\alpha \beta / 2)$. (This last formula agrees with $\theta_K(t)$ for the exactly solvable case in Eq.~\eqref{eq:alphaK} for times larger than the microscopic time scale: $t \gg \alpha^{-1}$). 
 Thus, Krylov winding is a direct consequence of the operator growth hypothesis~\cite{hyp}.

Since the Krylov index $n$ is interpreted as the position on the Krylov chain along which the operator travels~\cite{hyp}, the phase $e^{i \theta n}$ leads to a peak in the ``momentum-space Krylov wavefunction'' defined as:
\begin{equation}
C_K(t,\mu) \equiv \sum_{n \geq 0}\varphi_{n}^{2}(t_\beta)e^{i\mu n}
\end{equation}
with $\mu$ the momentum~\footnote{We note that this interpretation is ``dual'' to the one in holography, where size $l$ is regarded as momentum and $\mu$ gives the position.}.
For the solvable model of Eq.~\eqref{eq:ExactLinearWavefunction}, this gives 
\begin{equation}
\begin{aligned}
C_{K}(t,\mu) &= \frac{1}{\left(1-e^{i\mu}\tanh^2(\alpha t_\beta)\right)^{2\Delta}\cosh(\alpha t_\beta)^{4\Delta}}
\end{aligned}
\end{equation}
which has a peak located at momentum 
\begin{equation}\label{eq:muKThermoLimit}
\mu_K(t) = - 2 \theta_K(t) = -2\text{Arg}\left[\tanh(\alpha t_\beta)\right],
\end{equation}
whose width $\Delta \mu$ decays exponentially with time. (One can see the peak sharpening with time in Fig.~\ref{fig:Krylov_FTs}(a)). This implies that as an operator scrambles, its Krylov wavefunction becomes increasingly localized in momentum space, a phenomenon which is very generic and reveals a hidden coherence in the scrambling process.

\begin{figure}[h!]
\centering
  \includegraphics[width=0.8\columnwidth]
 {{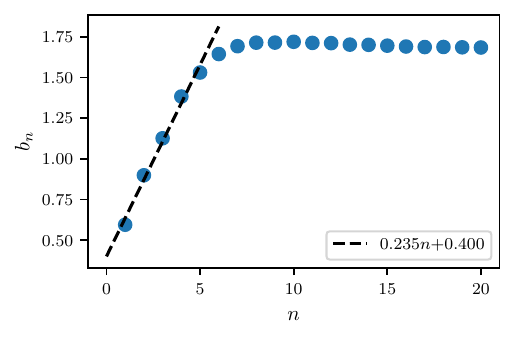}}
	\caption{Disorder-averaged Lanczos coefficients of the nonlocal spin Hamiltonian~\eqref{eq:NonlocalHamiltonian} in the main text with $N=8, \beta=1$, and $\mO=S_{1}^{x}$. The Lanczos coefficients are approximately described by a linear ramp at small $n$; a linear numerical fit is shown (dashed line). The coefficients saturate to a plateau which is approximately constant for $n\ll 4^{N}$. }
\label{fig:LanczosCoefficientsEndMatter}
\end{figure}

We have confirmed the presence of Krylov winding in two microscopic models.
First, in large-$q$ SYK, the Krylov wavefunction follows up to $1/q$ corrections the analytically solvable case of Eq.~\eqref{eq:ExactLinearWavefunction} (see SM Section \ref{app:SYK} \cite{supp}).
Second, we have also confirmed the presence of Krylov winding numerically in an anisotropic Heisenberg analog of the Sherrington-Kirkpatrick model~\cite{SherringtonKirkpatrick1975} of $N$ spin-$1/2$ degrees of freedom with all-to-all 2-local coupling:
\begin{equation}\label{eq:NonlocalHamiltonian}
H=\sum_{i<j}\sum_{\alpha=x,y,z}J_{ij}^{\alpha}S_{i}^{\alpha}S_{j}^{\alpha},
\end{equation}
where $J_{ij}^{\alpha}$ are drawn from a Gaussian distribution with mean 0 and variance $1/9N$~\cite{QuenchVsAnnealed}. 
We have first confirmed numerically that this model obeys the operator-growth hypothesis:
For $\mO = S_{1}^{x}$ at inverse temperature $\beta=1$, we constructed the Krylov basis generated by the seed operator 
$\rho_{\beta}^{1/4}\,\mO\,\rho_{\beta}^{1/4}$ and verified that the Lanczos coefficients $b_n$ grow linearly for $n\lesssim N$
(see Fig.~\ref{fig:LanczosCoefficientsEndMatter}, End Matter). 
We then computed $C_K(\mu,t)$ and observed a peak that sharpens as time increases (see Fig.~\ref{fig:Krylov_FTs}(b)). 
The corresponding peak position $\mu_K(t)$, shown in Fig.~\ref{fig:Krylov_FTs}(c), agrees with the analytic prediction in 
Eq.~\eqref{eq:muKThermoLimit} throughout the growth regime $t<\alpha^{-1}\log N$, using $\alpha\simeq 0.235$ extracted 
independently from the $b_n$. (See the End Matter for a discussion of late-time effects.)

\lsec{Relating size and Krylov winding}
Having established that Krylov winding is generic, we will derive sufficient conditions for it to generate size winding. The size winding ansatz \eqref{eq:SizeWindingDef} makes two assertions: (i) the phases of coefficients $c_P$ for all Pauli strings $P$ with the same size $|P|=\ell$ are aligned, and (ii) this common phase is linear in $\ell$. These properties are encoded in the size distribution $p(\ell,t) = \sum_{|P|=\ell} |c_{P}(t)|^{2}$ and the size-winding distribution $q(\ell,t) = \sum_{|P|=\ell} c_{P}^{2}(t)$. Perfect phase alignment corresponds to $|q(\ell,t)| = p(\ell,t)$, while phase linearity means $\arg(q(\ell,t)) \propto \ell$. These properties lead to a peak in the Fourier transform of the size winding distribution
\bea
C_S(\mu,t) \equiv   \sum_{\ell \geq 0} q(l) e^{i \mu \ell}
\eea
which is related to the fidelity of size winding-based quantum teleportation protocols~\cite{brown2023quantum,nezami2023quantum}. We note that $C_S$ depends on both the phase and the magnitude of the operator wavefunction $c_P(t)$: a peak in $C_S(\mu)$ requires the phase $\phi_P(t)$ to be coherent over a range of sizes for which the wavefunction amplitude is non-negligible.

By comparing the Fourier transform of the operator wavefunction in Krylov space (Fig. \ref{fig:Krylov_FTs}, $C_K(\mu,t)$) with its counterpart in size space (Fig. \ref{fig:size_FTs}, $C_S(\mu,t) $), it becomes clear that Krylov and size winding have similar phenomenology, and we now discuss their connection in more details.

To connect the two bases, we expand the operator in both:
\begin{equation}\label{eq:DecompositionEquivalence}
\opket{\rho_{\beta}^{1/2}\mO(t)} = \sum_{P}c_{P}(t)\opket{P} = \sum_{n} \varphi_{n}(t_\beta)\opket{\mO_{n}}.
\end{equation}
The size-winding distribution can be expressed in the Krylov basis as
\begin{equation}\label{eq:WindingDistKrylov}
q(\ell,t) = \sum_{n,m}\varphi_{n}(t_\beta)\varphi_{m}(t_\beta)M_{nm}(\ell),
\end{equation}
where $M_{nm}(\ell) \equiv \opbra{\mO_{n}}\hat{P}_{\ell}\opket{\mO_{m}}$ is the size-resolved Krylov overlap matrix (introduced in Ref.~\cite{Dissectingquantummanybodychaos}), with $\hat{P}_{\ell}$ the projector onto the subspace of size-$\ell$ Pauli strings.
Diagonalizing the $M$ matrix, we find $M_{nm}(\ell) = \sum_\nu \lambda_\nu(\ell) \psi_{\nu,n}(\ell) \psi_{\nu,m}(\ell)$ with $\psi_{\nu,n} \in \mathbb{R}$ the eigenvectors and $0 \leq \lambda_\nu \leq 1$ the eigenvalues \footnote{This is easily shown using the interlacing theorem since $M_{nm}$ is a submatrix of the matrix representation of a projector.}, which gives 
\begin{equation}\label{eq:sizeandwinding_eigen}
\begin{aligned}
p(\ell,t) =  \sum_\nu \lambda_\nu |Q_\nu(\ell,t)|^2, q(\ell,t) =  \sum_\nu \lambda_\nu Q_\nu(\ell,t)^2
\end{aligned}
\end{equation}
with $Q_\nu(\ell,t) = \sum_n \varphi_n(t_\beta) \psi_{\nu,n}(\ell)$.
Eq.~\eqref{eq:sizeandwinding_eigen} will now allow us to discuss phase alignment and linearity.

\emph{Phase alignment} is guaranteed if, for each size $\ell$, the matrix $M_{nm}(\ell)$ is rank-one, in which case $\lambda_0=1, \lambda_{\nu > 0}=0$ and $q(\ell,t) = Q_0(\ell,t)^2, p(\ell,t)=|Q_0(\ell,t)|^2$. In this case, there is only one state within each size sector that couples to the Krylov basis, ensuring that all contributions to $q(\ell,t)$ acquire the same phase from the Krylov wavefunction. This occurs in the SYK model in the large-$q$ limit as shown in Ref.~\cite{Dissectingquantummanybodychaos} (see also SM Section \ref{app:SYK} \cite{supp}). The interpretation is that, for large-$q$, operator growth generated by the Liouvillian is unidirectional: back-propagation to smaller sizes is suppressed~\cite{XiangyuSizeConcentration}. More generally, approximate phase alignment is expected if $M_{nm}(\ell)$ has a large spectral gap: $\lambda_0 \gg \lambda_{\nu>0}$. We conjecture that for $q$-local models, the spectral gap is parametrically large in $q$: $\lambda_0 \sim 1, \lambda_{\nu>0} \sim 1/q^\delta$ with some power $\delta>0$. 
Numerically, we have indeed observed that the $q=2$-local spin model of Eq.~\ref{eq:NonlocalHamiltonian} only has approximate phase alignment.

\emph{Phase linearity} depends on the relationship between the growth of Krylov complexity and operator size. The average Krylov index, or K-complexity, grows as $\langle n \rangle \sim e^{2\alpha t}$, while the average size grows as $\langle \ell \rangle \sim e^{\lambda_L t}$, where $\lambda_L$ is the quantum Lyapunov exponent. These rates are constrained by the chaos-operator growth (COG) bound $\lambda_L \leq 2\alpha$~\cite{hyp} which states that size cannot grow faster than K-complexity. Following Ref.~\cite{Dissectingquantummanybodychaos}, we define the ratio $h = \lambda_L / 2\alpha \in [0,1]$ which measures the saturation of this bound.

Let us now connect the phase of the Krylov wavefunction $\varphi_n$ with the phase of the winding distribution $q(l,t)$.
Assuming the rank-one condition, we have $q(\ell, t) = \left(\sum_n \varphi_n(t_\beta) \psi_{0,n}(\ell)\right)^2$.
As discussed in the SM Section \ref{app:phaselinearity}~\cite{supp}, the sum over $n$ is generically expected to be dominated by terms peaked around $n=n_0(\ell) \sim \ell^{1/h}$. Since $\psi_{0,n}$ is real, we predict
\begin{equation}\label{argumentsuperlinear}
\mathrm{Arg}[q(\ell,t)] \simeq \mathrm{Arg}[\varphi_{n_0(l)}^2(t_\beta)] \propto n_0(\ell) \propto \ell^{1/h}.
\end{equation}
This implies that size winding is linear in $\ell$ only when $h=1$, i.e., for systems that saturate the COG bound of \cite{hyp}. Since high-fidelity teleportation relies on a linear phase \cite{brown2023quantum, nezami2023quantum}, bound-saturating systems are optimal for such protocols.

To illustrate this, we use the SYK+bath model of Ref.~\cite{ChenSYKplusSYKBath,Dissectingquantummanybodychaos}, where the parameter $h$ can be tuned continuously. In this model, phase alignment is perfect and, using scramblon effective field theory \cite{KitaevSoftMode, GuKitaev, GuKitaevZhangTwoWayOTOC, ZhangLargeNMajoranaSizeDist, Zhang_2021, Zhang_2023, Liu_2024} and results from Ref.~\cite{Dissectingquantummanybodychaos}, we find that the size-winding distribution takes the form of a ``compressed exponential'':
\begin{equation}\label{Eq:qoflmain}
    q(\ell,t)  \propto \exp\left(- K^{-1/h}(\ell-\ell_0)^{1/h} e^{-2\alpha (t+i\beta/4)}\right),
\end{equation}
where $K$ is a constant and $\ell_0$ is the initial average size. (See SM \cite{supp} for the full expression of $q(\ell,t)$, along with derivations, and for a plot of $q(\ell,t)$ for various values of $h$).
Eq.~\eqref{Eq:qoflmain} should be contrasted with the Krylov wavefunction in Eq.~\eqref{eq:varphilatetimes} which is exponential in $n$ (see Fig.~\ref{fig:Schematic} for a schematic comparison).

As predicted in Eq.~\eqref{argumentsuperlinear}, the phase of the winding distribution thus scales as $\ell^{1/h}$ in this model. For $h<1$, the superlinear phase winding broadens the peak of the Fourier transform $C_S(\mu,t) = \sum_\ell q(\ell,t) e^{i\mu\ell}$, see Fig.~\ref{fig:size_FTs}, and qualitatively changes the shape of the peak, with e.g. an ``elbow'' appearing to the left of the peak, see Fig. \ref{fig:size_FTs}(b) at $h=0.5$ for an example. Since the peak in $\mu$ is interpreted as a well-defined position of an infalling particle in holography, it would be interesting to study the holographic interpretation of such non-standard behavior in $C_S(\mu,t)$ for models that do not saturate the COG bound.

Before moving to the conclusion, let us comment on the dependence on the initial operator $O$.
Krylov winding is expected for any generic local operator, with the dependence on the initial operator entering primarily through the quantitative value of the asymptotic slope $\alpha$ of the Lanczos coefficients, which can vary with $O$.
By contrast, \emph{size winding} relies on additional conditions---an approximately low-rank Krylov-to-size map and the degree of saturation of the chaos--operator-growth bound $\lambda_L \le 2\alpha$.
We expect these conditions to be satisfied to varying degrees for different $O$s, and hence the features of size winding to vary from operator to operator (e.g.\ both the spectral gap controlling the low-rank approximation and the ratio $h=\lambda_L/(2\alpha)$ may have such a dependence).

In conclusion, we have shown that Krylov winding provides a generic mechanism for the emergence of coherence in many-body quantum dynamics and is a direct consequence of the operator growth hypothesis.
Krylov winding should also appear in few-body (semi-)classical systems~\cite{bohigas1984spectral,PhysRevB.100.155128}, since the operator growth hypothesis applies there as well~\cite{hyp}, including classically integrable models with unstable saddles~\cite{xu2020does, xiangyusaddledom}.
It would be interesting to develop a classical phase-space interpretation of Krylov winding in such settings, and to ask whether analogs of \emph{size} winding exist.

\lsec{Acknowledgments} This work was supported by the U.S. Department of Energy, Office of Science, Office of Basic Energy Sciences under Early Career Research Program Award Number DE-SC0025568. We acknowledge the Digital Research Alliance of Canada for computational resources. We thank Ehud Altman and Xiangyu Cao for invaluable contributions to the manuscript and Pawel Caputa and Tommy Schuster for illuminating discussions.

\lsec{Data availability} The data and simulation codes that
support the findings of this article are openly
available \cite{dataGitHub}.
\bibliography{refs}

\onecolumngrid
\begin{center}
\ \vskip 0.2cm
{\large\bf End Matter}
\end{center}
\twocolumngrid

\renewcommand{\theequation}{A\arabic{equation}}
\stepcounter{superequation}

\emph{Late times and finite-size effects.}---
In the main text we have only focused on the growth regime $t \ll \alpha^{-1}\log(N)$ for which finite-size effects are negligible since the average operator size and Krylov index are much smaller than the system size $N$. 
However, late-time, finite-size effects in size winding show rich behavior, as discussed in Refs.~\cite{ZhouSizeWinding,schuster2022many,nezami2023quantum,gao2024commuting}, and can indeed be observed in our finite-size numerics for the spin model (See Figs. \ref{fig:Krylov_FTs}b and \ref{fig:size_FTs}c).

It is interesting to contrast finite-size effects in the Krylov and size bases.
Size can only take $N$ different values, leading to a depth for the basis of Pauli strings that is capped at $\ell = N$, and a saturation of the average size to a value of order $N/2$ after the scrambling time $\lambda_L^{-1} \log(N)$~\cite{Roberts_2017,ZhangLargeNMajoranaSizeDist, ZhouSizeWinding}. In the late-time regime, the peak in $C_S(\mu,t)$ approaches $\mu = 0$ (see Fig.~\ref{fig:size_FTs}c), signaling the fact that the operator wavefunction stops moving in $\ell$ space and thus has a vanishing average ``momentum'' $\mu$.

\begin{figure}[t!]
\centering
  \includegraphics[width=0.9\columnwidth]
 {{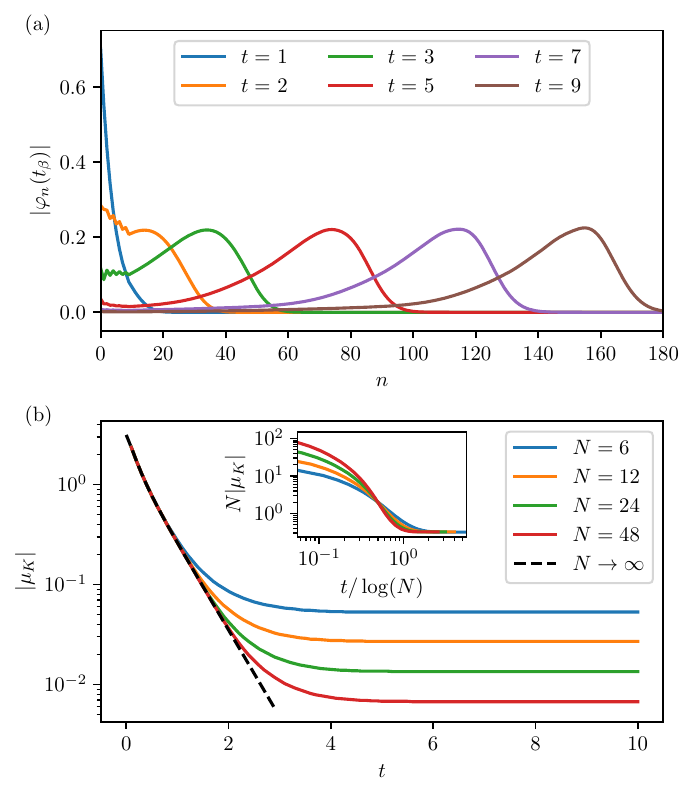}}
	\caption{Numerically-obtained results for the ramp-plateau model with $\beta=\alpha=1$. (a) Krylov wavefunctions for $N=10$. For $t \gg \alpha^{-1} \log N$, the Krylov wavefunction approximately retains its shape and propagates ballistically. (b) The (absolute value of the) location of the peak in $C_K(\mu,t)$, as a function of $t$ and $N$. For $t \gg \alpha^{-1} \log N$, $\mu_K$ approaches a non-zero value due to finite size effects. The $N\to\infty$ result is given by Eq.~\eqref{eq:muKThermoLimit}. Inset: The rescaled winding peak, $N|\mu_K|$, vs. $t/\log(N)$. This rescaling collapses the scrambling time and late time plateaus.}
\label{fig:KrylovWavefunctionsEndMatter}
\end{figure}

The story is different in the Krylov basis.
There, the Krylov chain has a length that is \emph{exponential} in $N$ and the average value of the Krylov index can thus continue increasing past $n \sim N$. 
Finite-size effects do have an effect on the Krylov basis however: as shown in Fig.~\ref{fig:LanczosCoefficientsEndMatter}, the Lanczos coefficients $b_n$ grow linearly for $n< \mathcal{O}(N)$ and saturate to a roughly constant value for $n>\mathcal{O}(N)$.
In SM Section \ref{app:RampPlateau} \cite{supp}, we introduce a simple ramp-plateau model in order to study finite-size effects in the Krylov basis:
\begin{equation}\label{eq:RampPlateauDef}
b_{n} = \begin{cases}
    \alpha n, & n\leq N\\
    \alpha L, & n>N
\end{cases}
\end{equation}
The exponential spreading of the Krylov wavefunction in the ramp region for $t \ll \alpha^{-1} \log(N)$ transitions into ballistic propagation of a wavepacket in the plateau region for $t \gg \alpha^{-1} \log(N)$ (See Fig.~\ref{fig:KrylovWavefunctionsEndMatter}). This is natural since, for constant $b_n$, the Krylov dynamics of Eq.~\ref{hoppingmodel} describes a discrete advection equation with constant velocity.
The ballistic propagation of the operator wavefunction translates into an average momentum $\mu$ that remains at a constant non-zero value $\mu_K$ in that regime, as observed in Fig.~\ref{fig:KrylovWavefunctionsEndMatter}(b) for the ramp-plateau model, and in Fig.~\ref{fig:Krylov_FTs}(c) for our finite-size numerics of the spin model.

\clearpage

\begin{widetext}
\begin{center}
\textbf{\large Supplemental Material for: \\ \titlename}
\end{center}
\end{widetext}
\setcounter{equation}{0}
\setcounter{figure}{0}
\setcounter{table}{0}
\makeatletter
\setcounter{secnumdepth}{2}
\setcounter{section}{0}
\setcounter{subsection}{0}
\renewcommand{\thesection}{\Alph{section}}
\renewcommand{\thesubsection}{\Roman{subsection}}
\renewcommand{\theequation}{S\arabic{equation}}
\renewcommand{\thefigure}{S\arabic{figure}}
\renewcommand{\bibnumfmt}[1]{[#1]}
\renewcommand{\citenumfont}[1]{#1}
\renewcommand{\theHfigure}{S\arabic{figure}}
\section{Large-$q$ SYK coupled to bath}\label{app:SYK}
In this section, we review the model of Ref. \cite{ChenSYKplusSYKBath, Dissectingquantummanybodychaos} with Majorana SYK$_4$ ($N$ sites) coupled to a Majorana SYK$_4$ bath ($N^2$ sites), which allows us to tune $h = \lambda_L/2\alpha$ continuously. The Hamiltonian of this model is given by 
\begin{equation} \label{Eq:SYK+SYKBathHam}
    \mathcal{H} = \mathcal{H}_\chi + \mathcal{H}_{\psi} + \mathcal{H}_c
\end{equation}
with 
\begin{subequations}
\begin{align}
    \mathcal{H}_\chi &= \frac{1}{4!}  \sum_{i,j,k,l=1}^N J_{ijkl} \chi_i \chi_j \chi_k \chi_l \\
    \mathcal{H}_\psi &= \frac{1}{4!} \sum_{a,b,c,d=1}^{N^2} J'_{abcd} \psi_a \psi_b \psi_c \psi_d \\
    \mathcal{H}_c &= \frac{1}{(2!)^2} \sum_{i,j=1}^N\sum_{a,b=1}^{N^2} u_{ijab} \chi_i \chi_j \psi_a \psi_b
\end{align}
\end{subequations}
where the couplings $\{J_{ijkl}\}, \{J'_{abcd}\}, \{u_{ijab}\}$ are independent Gaussian random numbers with zero mean and variance, 
\begin{equation}
    \overline{J^2_{ijkl}} = \frac{3!J^2}{N^3}, \overline{J'^2_{abcd}} = \frac{3!J^2}{N^6}, \overline{u^2_{ijab}} = \frac{2!u^2}{N^5}.
\end{equation}
In the strong-coupling limit (or low temperature) $\beta J \gg 1$, the Schwinger-Dyson equations are analytically solvable. In this limit, the auto-correlation function for the $\chi$ fermions is given by 
\begin{equation}
    C_{\chi}(t) \propto \left(\cosh \frac{\pi t}{\beta}\right)^{-1/2},
\end{equation}
which is the same as the usual SYK model without any bath coupling and the Lyapunov exponent for the OTOC $F_{\chi \chi}(t_1, t_2)$ between $\chi$ fermions is given by 
\begin{equation} \label{eq:lambdaSYKbath}
    \lambda_L = \frac{2\pi}{\beta}\left(1 - \frac{\sqrt{k^4 +4k^2}-k^2}{2}\right)
\end{equation}
with $k = u^2/J^2$ \cite{ChenSYKplusSYKBath}. 

This model admits a natural generalization to $q$-body interactions and is solvable in the large-$q$ limit. We assume that the tunability of the Lyapunov exponent due to coupling with the bath (Eq.~\eqref{eq:lambdaSYKbath}) persists in this limit.
\subsection{Krylov wavefunction}
In this section, we obtain the Krylov wavefunction and its Fourier transform for the operator $\rho_\beta^{1/2}\chi(t)$. At large $q$ as well, the two-point function of the SYK+bath model at strong coupling is the same as the usual SYK model. Consequently, the Lanczos coefficients are given by those of the large-$q$ SYK model \cite{hyp},
\begin{equation}
    b_n =
    \begin{cases}
        \nu \frac{\pi}{\beta} \sqrt{\frac{2}{q}} + O\left(\frac{1}{q}\right), & \text{if } n = 1 \\
        \nu \frac{\pi}{\beta} \sqrt{n(n-1)} + O\left(\frac{1}{q}\right), & \text{if } n > 1 
    \end{cases}
\end{equation}
which satisfies the operator growth hypothesis \cite{hyp} $b_n \overset{n \gg 1}{\rightarrow} \alpha n$ with $\alpha = \pi\nu/\beta$. Here, $\beta \mathcal{J} = \pi \nu/\cos(\pi\nu/2) $. The Krylov wavefunction for this model is given by 
\begin{equation}
    \varphi_n(t) =
    \begin{cases}
        1 + \frac{2}{q}\ln \mathrm{sech} \alpha t +  O(\frac{1}{q^2}), & \text{if } n = 0 \\
        (\tanh \alpha t)^n \sqrt{\frac{2}{nq}}+ O(\frac{1}{q^2}), & \text{if } n > 0. 
    \end{cases}
\end{equation}
As explained in the main text, the Krylov wavefunction for the operator $\rho_\beta^{1/2} \chi(t)$ is obtained by replacing $t \rightarrow t_\beta \equiv t+ i \beta/4$ in the above wavefunction. The Fourier transform $C_K(t_\beta, \mu)$ is then computed exactly:
\begin{subequations}
    \begin{align}
        C_{K}(t,\mu) &\equiv \sum_{n=0}^{\infty}\varphi_{n}(t_\beta)^2e^{i n \mu}\\
&= 1 + \frac{2}{q} \ln \left(\frac{1 - \tanh^2 \alpha t_\beta}{1 - e^{i \mu}\tanh^2 \alpha t_\beta }\right)  + O\left(\frac{1}{q^2}\right).
    \end{align}
\end{subequations} 
Note that the Krylov wavefunction $\varphi_n(t_\beta)$ and the Fourier transform $C_K(t, \mu)$ matches with those of the analytically solvable case of Eq. \ref{eq:ExactLinearWavefunction} in the main text up to $1/q$.

\subsection{Size and winding distribution in Majorana basis}
In this section, we obtain the size and winding distribution of the operator $ \rho_\beta^{1/2}\chi(t)$ in the SYK+bath model [Eq. \ref{Eq:SYK+SYKBathHam}]. Any operator can be expressed in the Majorana basis as $\hat{O}(t)= \sum_\ell \sum_{j_1 < j_2 < \dots < j_\ell} i^{[n/2]} c_{j_1 j_2 \dots j_\ell}(t) \chi_{j_1} \chi_{j_2} \dots \chi_{j_\ell}$ with the convention $\{\chi_j, \chi_k\} = 2 \delta_{jk} $. The \emph{size} of the basis operator is defined as the number of non-identity operators in the string. For example, the size of the operator $\chi_{j_1} \chi_{j_2} \dots \chi_{j_\ell}$ is $\ell$. Next, the size distribution $P(\ell,t)$ and the size winding distribution $Q(\ell, t)$ are defined as 
\begin{align}
    P(\ell, t) &= \sum_{j_1 <j_2 < \dots <j_\ell} |c_{j_1 j_2 \dots j_\ell}(t)|^2 \\
    Q(\ell, t) &= \sum_{j_1 <j_2 < \dots <j_\ell} (c_{j_1 j_2 \dots j_\ell}(t))^2 
\end{align}
In the thermodynamic limit ($N \to \infty$), it is convenient to define normalized size as $s \equiv \ell/N$ which becomes a continuous variable in the range $[0,1]$ and the normalized distribution functions as $p(s, t) = N P(sN, t), q(s, t) = N Q(sN, t) $.
Ref. \cite{ZhouSizeWinding} showed that the size and winding distribution of generic chaotic large-$N$ quantum systems with all-to-all interactions can be obtained using scramblon effective field theory \cite{KitaevSoftMode, GuKitaev, GuKitaevZhangTwoWayOTOC, ZhangLargeNMajoranaSizeDist, Zhang_2021, Zhang_2023, Liu_2024},
\begin{multline}\label{Eq:ScramblonPQ}
    p/q(s, t) \approx \int_0^\infty dy \frac{h^R(y, T_{12})}{\sqrt{2 \pi \sigma^2}} \times\\ \exp\left(-\frac{1}{2\sigma^2}\left(s - \frac{1-f^A(\lambda y, -i\beta/2)}{2}\right)^2\right)
\end{multline}
where $\sigma^2 = (1- f^A(\lambda y, -i\beta/2)^2)/4N$, $T_1 = t - i \epsilon, T_3 = 0, T_4 = i \beta/2$,  $\lambda$ is the scramblon propagator, and $h^R$ and $f^A$ are functions defined below. For the size distribution $p(s,t)$, $T_2 = t$, $\lambda = C^{-1} \exp(\lambda_L t) \equiv \lambda_0$. For the winding distribution $q(s,t)$, $T_2 = t+i\beta/2$, $\lambda = C^{-1} \exp(\lambda_L t) \exp(i\lambda_L \beta/4)$.

Following \cite{GuKitaevZhangTwoWayOTOC, ZhouSizeWinding}, the vertex functions $\Upsilon^{R/A,m}$ are expressed as moments of $h^{R/A}$ 
\begin{equation}
    \Upsilon^{R/A,\sum_l m_l}(T_{12}) = \int_0^\infty dy y^{\sum_l m_l} h^{R/A}(y, T_{12}) 
\end{equation}
and introduce another function $f^A$ resulting from summing over the scramblon modes 
\begin{align}
    &\sum_l \frac{(-\lambda y)^{m_l}}{m_l!} \Upsilon^{A, m_l}(T_{34})  \nonumber \\
    &= \int_0^\infty dy_l e^{-\lambda y y_l} h^A(y_l, T_{34}) \equiv f^A(\lambda y, T_{34}).
\end{align}

For the usual large-$q$ SYK model, the vertex functions $\Upsilon^{R/A, m}(T)$ and the corresponding functions $f^A(\lambda y, T), h^A(y, T)$ are given by \cite{GuKitaevZhangTwoWayOTOC, ZhouSizeWinding}
\begin{align}
    h^R(y,T_{12}) &= \frac{y^{2 \Delta -1} \cos ^{2 \Delta }\left(\frac{\pi  v}{2}\right) }{\Gamma (2 \Delta )} \nonumber \\
    & \times\exp \left(- y \cos \left( \pi v \left(\frac{1}{2}- \frac{i T_{12}}{\beta}\right)  \right) \right) \label{eq:hLargeqSYK}\\
  f^A(\lambda y, T_{34}) &= \cos ^{2 \Delta }\left(\frac{\pi  v}{2}\right) \nonumber \\
  &\times \left(
  \cos \left( \pi v \left(\frac{1}{2}- \frac{i T_{34}}{\beta}\right)  \right)
  +\lambda  y\right)^{-2 \Delta}.
\end{align}

Following the prescription in Ref.~\cite{Dissectingquantummanybodychaos} (see also \cite{Maldacena_2016, shenker2015stringyeffectsscrambling}), we assume that the vertex functions are modified for $h<1$ such that the following relation holds: 
\begin{equation}
    \tilde{h}^{R/A}(y, T_{ij}) = \frac{1}{h}y^{1/h - 1} h^{R/A}(y^{1/h}, T_{ij}).
\end{equation}
Using this, the size and winding distributions are modified as  
\begin{multline}\label{Eq:ScramblonPQModified}
    p/q(s, t) \approx \int_0^\infty dy \frac{h^R(y, T_{12})}{\sqrt{2 \pi \tilde{\sigma}^2}} \\ \exp\left(-\frac{1}{2\tilde{\sigma}^2}\left(s - \frac{1-\tilde{f}^A(\lambda y^h, -i\beta/2)}{2}\right)^2\right)
\end{multline}
where $\tilde{\sigma}^2 = (1- \tilde{f}^A(\lambda y^h, -i\beta/2)^2)/4N$ and
\begin{equation}
    \tilde{f}^A(\lambda y, T_{34}) = \int_0^\infty dy_{l} e^{-\lambda y y_l^h} h^A(y_l, T_{34}).
\end{equation}

In the early-time regime ($\lambda_0 \ll 1$),  we can approximate the Gaussian in Eq. \eqref{Eq:ScramblonPQModified} as a Dirac delta function, to get  
\begin{subequations}
    \begin{align}
        p(s,t) &= 2 |\partial_y \tilde{f}^A(\lambda_0 y^h, -i \beta/2)|^{-1}h^R(y,0)\\
        q(s,t) &= 2 |\partial_y \tilde{f}^A(\lambda_0 y^h, -i \beta/2)|^{-1} e^{-i \pi \nu /2} \nonumber \\
        &~~~~~~~~~~~~~ \times h^R(e^{-i \pi \nu /2}y,-i\beta/2)
    \end{align}
\end{subequations}
where $1-2s = \tilde{f}^A(\lambda_0 y^h, -i \beta/2)$. Substituting Eq. \ref{eq:hLargeqSYK} above yields,
\begin{align}
    \mathrm{Arg}[q(s(y),t)] &= y \sin{(\pi \nu/2)} - \pi \nu \Delta, \\
    p(s(y),t) &= \frac{(\lambda_0 h)^{-1} 2 y^{2 \Delta - h} e^{-y \cos(\pi \nu/2)}}{ \int_0^\infty dy_l y_l^{2 \Delta + h - 1} e^{-\lambda_0 (y y_l)^h - y_l}},\label{Eqs:SYK+SYKDistr} \\
    |q(s(y),t)| &= p_S(s(y), t)
\end{align}
with 
\begin{equation}\label{Eq:sofy}
    1 - 2s(y)= \frac{\cos^{2\Delta}(\pi \nu/2)}{\Gamma(2 \Delta)}\int_0^\infty dy_l y_l^{2 \Delta - 1} e^{-\lambda_0 (y y_l)^h - y_l}.
\end{equation}

From the above expression, note that the size $s$ is a monotonic function of $y$ in the range $[s_0, 1/2]$ with $s_0 \equiv (1-\cos^{2\Delta}(\pi \nu/2))/2$. Here, $Ns_0$ is the initial average size of the operator $\rho_\beta^{1/2}\chi$.
\begin{figure}[t!]
	\centering	
  \includegraphics[width=0.8\columnwidth]
{{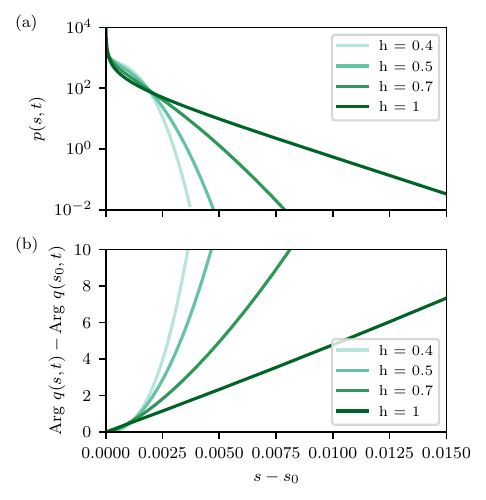}}
	\caption{Results for the large-$q$ SYK+bath model~\cite{ChenSYKplusSYKBath,Dissectingquantummanybodychaos} at $N=3000, t=0.9/2\alpha, \nu=0.5, q=6$: (a) Size distribution $p(s, t)$ and (b) winding phase $\mathrm{Arg}~q(s, t)$ for different values of $h=\lambda_L/2\alpha$. Here $s=\ell/N$. The phase is linear only for the bound-saturating case $h=1$.}
\label{fig:SYK}
\end{figure}

For $h<1$, the integrals in Eqs. (\ref{Eqs:SYK+SYKDistr}), (\ref{Eq:sofy}) cannot be computed analytically. We compute the integrals numerically to obtain the size distribution and the phase of the winding distribution in Fig. \ref{fig:SYK}. In the early time regime ($\lambda_0 \ll 1$), the integrals can be computed by approximating the exponential to linear order $e^{-\lambda_0 (y y_l)^h} \approx 1 -\lambda_0 (y y_l)^h $ to get,  
\begin{align}\label{Eq:sofyLinear}
    y &= K^{-1/h} e^{-2 \alpha t} (s-s_0)^{1/h}, \\ \label{Argq}
    \mathrm{Arg} (q(s,t))  &= \sin{(\pi \nu/2)} K^{-1/h} e^{-2 \alpha t} (s-s_0)^{1/h}  - \pi \nu \Delta, 
\end{align}
and
\begin{multline}
    p(s,t) = 
         \frac{8N\Delta^2 \cos(\pi \nu/2)}{h \Gamma(2 \Delta +h)} K^{-2\Delta/h +1}   (s-s_0)^{2\Delta/h -1} \\ \times e^{-4 \Delta \alpha t} \exp(-K^{-1/h} \cos(\pi \nu/2) e^{-2 \alpha t} (s-s_0)^{1/h}),
\end{multline}
where $K = \cos^{2\Delta-1}(\pi \nu/2)\Gamma(2\Delta +h)/{4N\Delta\Gamma(2\Delta+1)}.$ 
We thus see explicitly from Eq.~\eqref{Argq} that the phase of $q$ winds superlinearly in general: $\mathrm{Arg} (q(s,t)) \sim (s-s_0)^{1/h}$, with $ h \leq 1$.
(The value of $C$, computed using the ladder identity \cite{GuKitaev} and the vertex functions, could depend on the details of the model \ref{Eq:SYK+SYKBathHam}. Since we do not explicitly compute the vertex functions of the model \ref{Eq:SYK+SYKBathHam}, we take the value of large-$q$ SYK model $C = 4 N \Delta^2 \cos(\pi \nu/2)$ \cite{GuKitaevZhangTwoWayOTOC}.)

Next, we compute the Fourier transform of the size winding distribution
\begin{align}
    C_S(\mu, t) &\equiv \sum_\ell Q(\ell, t) e^{i \mu \ell} \\
    &= \int_{s_0}^{1/2} ds \ q(s,t ) e^{i \mu s N} \\ 
    &= \int_0^\infty dy \frac{ds(y)}{dy} p(y,t) e^{i \mathrm{Arg}q(y,t) + i \mu s(y) N} \\
    &= \frac{\cos^{2 \Delta}\pi \nu/2}{\Gamma(2 \Delta)} \int_0^\infty dy y^{2 \Delta - 1} e^{-y \cos \pi \nu/2} \nonumber \\
    &\ \ \ \ \ \ \ \ \ \ \ \ \ \ \ \times e^{i (y \sin(\pi \nu/2) - \pi \nu \Delta) + i \mu s(y)N}.\label{Eq:FTSizeWindDist}
\end{align}
Since we are interested in the early time regime, we take the first-order approximation to $s(y)$ of Eq. (\ref{Eq:sofyLinear}) to get 
\begin{align}
    C_S(\mu, t) &= \frac{\cos^{2 \Delta}\pi \nu/2}{\Gamma(2 \Delta)} e^{i (\mu s_0 N - \pi \nu \Delta)} \nonumber \\
    & \times\int_0^\infty dy y^{2 \Delta - 1}  e^{-y e^{-i \pi \nu/2} + i \mu K N  e^{2 \alpha h t}y^h}. 
\end{align}
We compute the above integral numerically for $h<1$ to obtain Fig. \ref{fig:size_FTs}(b) in the main text. For $h=1$, the above integral can be computed analytically, 
\begin{equation}
    C_S(\mu, t) = \left(\frac{\cos(\pi \nu/2)}{ e^{-i \pi \nu/2} - i \mu K N  e^{2 \alpha h t}}\right)^{2 \Delta} e^{i (\mu s_0 N - \pi \nu \Delta)}
\end{equation}
and is used in Fig. \ref{fig:size_FTs}(a) in the main text.
\subsection{\emph{Phase alignment}: Size-resolved Krylov overlap matrix }
In this section, following Ref.~\cite{Dissectingquantummanybodychaos}, we show that the size-resolved Krylov overlap matrix $M_{nm}(\ell)$ is rank-1. First, we calculate the size winding distribution at different times defined as $Q(\ell, t_1, t_2) \equiv  \sum_{j_1 <j_2 < \dots <j_\ell} c_{j_1 j_2 \dots j_\ell}(t_1)c_{j_1 j_2 \dots j_\ell}(t_2)$. In the thermodynamic limit, we define the normalised size winding distribution as $q(s, t_1,t_2) = N Q(sN, t_1, t_2)$. Proceeding as in the previous section gives us,
\begin{align}
    q(s,t_1, t_2) &= 2 |\partial_y \tilde{f}^A(\lambda_0 y^h, -i \beta/2)|^{-1} e^{-i\pi \nu/2} \nonumber \\
    & \ \ \ \ \ \times h^R(e^{-i\pi \nu/2}y,t_{12}-i\beta/2)
\end{align}

where $1-2s = \tilde{f}^A(\lambda_0 y^h, -i \beta/2)$, $\lambda_0 = C^{-1} \exp[\lambda_L (t_1 + t_2)/2], t_{12} = t_1 - t_2$. Substituting Eq. \ref{eq:hLargeqSYK} and and taking the linear approximation for $s(y)$ in the early time-regime $y = K^{-1/h} e^{- \alpha (t_1 + t_2)} (s-s_0)^{1/h}$, we get
\begin{align}
    &q(s,t_1, t_2) \nonumber \\
    &= \frac{8N\Delta^2 \cos(\pi \nu/2)}{h \Gamma(2 \Delta +h)} K^{-\frac{2\Delta}{h} +1}   (s-s_0)^{\frac{2\Delta}{h} -1} \nonumber \\
    &\times \exp[-i\pi\nu\Delta-2 \Delta \alpha (t_1+t_2)] \nonumber \\
         & \times\exp\left[-K^{-1/h}  (s-s_0)^{1/h}  \frac{(e^{-2\alpha t_2} + e^{-2\alpha t_1})}{2}e^{-i \pi \nu /2}\right] \nonumber 
\end{align}
which factorizes as $q(s, t_1, t_2) = r(s,t_1)r(s,t_2)$ with
\begin{align}
    &r(s,t)  \nonumber \\
    &=\sqrt{\frac{8N\Delta^2 \cos(\pi \nu/2)}{h \Gamma(2 \Delta +h)}} K^{-\Delta/h +1/2}   (s-s_0)^{\Delta/h -1/2} \nonumber \\ 
    &\times \exp\left[-2 \Delta \alpha t_\beta-(1/2)K^{-1/h}  (s-s_0)^{1/h} e^{-2 \alpha t_\beta}\right] \label{rofs}
\end{align}

Now, recall from the main text how to express $q$ in terms of the Krylov wavefunction: 
\bea
q(s,t_1,t_2) &=& N\sum_{nm} \varphi_n(t_{1,\beta}) \varphi_m(t_{2,\beta}) M_{nm}(\ell) 
\eea
with the size-resolved Krylov overlap matrix $M_{mn}(l) \equiv (O_n | P_l | O_m) $ and $t_\beta = t + i\beta/4$.
Since $q(s,t_1,t_2) = r(s,t_1)r(s,t_2)$, it implies that $M_{mn}(l) $ must be rank-1.

\subsection{\emph{Phase (super-)linearity}}\label{app:phaselinearity}
We already showed above that the phase of the winding distribution goes as $\mathrm{Arg} (q(s,t)) \sim (s-s_0)^{1/h}$, and is thus superlinear for $h < 1$.
By contrast, Krylov winding is always linear: $\mathrm{Arg}(\varphi_n) \sim n$.
Yet, $q(s,t)$ and $\varphi_n$ are describing the same operator in two different bases, so one should be able to connect the two different forms of winding.
In this section, we explain this connection.

The distributions $q(l,t)$ and $\varphi_n$ are connected through the relation: $q(l, t) = N \left( \sum_n \varphi_n(t_\beta) \psi_{0,n}(\ell) \right)^2$.
The key point is that this sum over $n$ is peaked around a value $n_0(l) \sim l^{1/h}$, as explained further below. This means that 
\begin{align}
    q(l, t) &=N \left( \sum_n \varphi_n(t_\beta) \psi_{0, n}(\ell)\right)^2 \nonumber\\
    &\approx N \left(\varphi_{n_0(\ell)}(t_\beta) \psi_{0, n_0(\ell)}(\ell)\right)^2
    \label{approxpeak}
\end{align}
Since $ \psi_{0, n}(\ell)$ is real, we have $\mathrm{Arg}~q(l) \sim \mathrm{Arg}~\varphi_{n_0(l)}^2 \sim \ell^{1/h}. $

 Let us now show numerically that the terms in the sum $\sum_n \varphi_n(t_\beta) \psi_{0,n}(\ell)$ are peaked around a value $n(l)$.
 First, we need to calculate $\psi_{0, n}(\ell) $, which we do as follows. 
 Using Eqs. \ref{eq:ExactLinearWavefunction} and \ref{rofs}, and defining $y = \tanh \alpha t_\beta$, the relation $\sum_n \varphi_n(t_\beta) \psi_{0,n}(\ell) = r(s,t)/\sqrt{N} $ derived above becomes
\begin{align}
    &\sum_n \sqrt{\frac{\Gamma(2\Delta+n)}{\Gamma(2\Delta)\Gamma(n+1)}} \psi_{0,n}(\ell)  y^n \nonumber \\
    &~~~= \sqrt{\frac{8\Delta^2 \cos(\pi \nu/2)}{h \Gamma(2 \Delta +h)}}    \left(\frac{s-s_0}{K}\right)^{\Delta/h -1/2} \nonumber \\ 
    &~~~~~\times\frac{1}{(1+y)^{2\Delta}} \exp\left[- \frac{1}{2}\left(\frac{s-s_0}{K}\right)^{1/h} \left(\frac{1-y}{1+y}\right) \right].
\end{align}
By performing a Taylor expansion of the RHS about $y=0$ and matching the coefficients with the LHS, we extract $\psi_{0,n}(\ell)$: 
\begin{align}
    &\psi_{0,n}(\ell) \nonumber \\
    &=\sqrt{\frac{8\Delta^2 \cos(\pi \nu/2)\Gamma(2\Delta)}{h\Gamma(2\Delta +h)}} e^{-\tilde{l}^{1/h}/2}\tilde{l}^{\Delta/h - 1/2} \nonumber \\
    &\times \sqrt{\Gamma(2\Delta+n)n!} \sum_{m=0}^n \frac{(-1)^{n-m} \tilde{l}^{m/h}}{m!(n-m)!\Gamma(2\Delta+m)} \\
    &= \sqrt{\frac{8\Delta^2 \cos(\pi \nu/2)}{h\Gamma(2\Delta +h)\Gamma(2 \Delta)}} e^{-\tilde{l}^{1/h}/2}\tilde{l}^{\Delta/h - 1/2} \nonumber \\
    &\times \sqrt{\frac{\Gamma(2\Delta+n)}{n!}} (-1)^n ~_1F_1(-n, 2\Delta, \tilde{l}^{1/h})
\end{align}
where we defined $\tilde{l} =(s-s_0)/K $ and $_1F_1$ is the confluent hypergeometric function. 

\begin{figure}[t]
	\centering	
  \includegraphics[width=0.99\columnwidth]
{{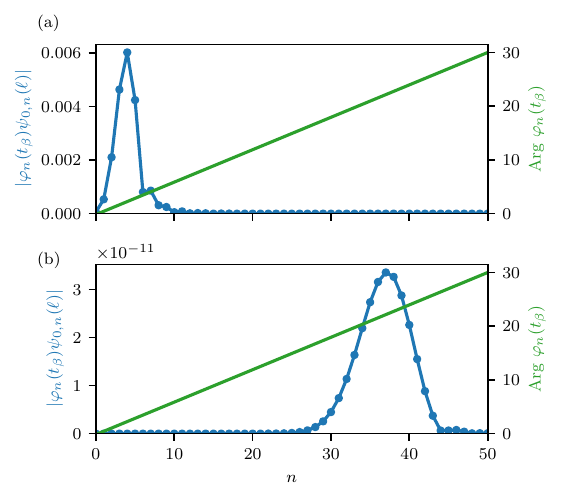}}
	\caption{Peak in $|\varphi_n(t_\beta)\psi_{0,n}(\ell)|$ vs $n$: (a) $h=1$ and (b) $h=0.5$ at $N= 3000, \nu = 0.5, \Delta = 1/6, t = 0.9/2\alpha, \ell = (s_0+0.01)N$. The green curves show that the phase $\mathrm{Arg}~\varphi_n(t_\beta)$ varies little across the peaks.}
\label{fig:Q0Peak}
\end{figure}

We can now study the terms in the sum as a function of $n$:
\begin{align}
    &|\varphi_n(t_\beta)|\psi_{0,n}(\ell) \nonumber \\
    &=\sqrt{\frac{8\Delta^2 \cos(\pi \nu/2)}{h\Gamma(2\Delta +h)}} \frac{e^{-\tilde{l}^{1/h}/2}\tilde{l}^{\Delta/h - 1/2}}{|\cosh\alpha t_\beta|^{2\Delta}} \nonumber \\
    &\times |\tanh \alpha t_\beta|^n \frac{(-1)^n}{\Gamma(2\Delta) \Gamma(n+1)} ~_1F_1(-n, 2\Delta, \tilde{l}^{1/h}).
\end{align}
As we show in Fig.~\ref{fig:Q0Peak}, these terms are peaked around a value $n(l)$, and the phase of $\varphi_n$ varies little within that peak, such that Eq.~\eqref{approxpeak} is a good approximation.

\section{Ramp-plateau model}\label{app:RampPlateau}

In this appendix, we analyze a toy model of operator dynamics for a finite chaotic system of size $N$. Our toy model treats the Lanczos coefficients by assuming exact linear growth up to a finite size scale beyond which the Lanczos coefficients are constant,
\begin{equation}\label{eq:RampPlateauDef}
b_{n} = \begin{cases}
    \alpha n, & n\leq N\\
    \alpha L, & n>N
\end{cases}
\end{equation}
This description is a qualitatively accurate model of nonlocal spin systems, such as the Hamiltonian~\eqref{eq:NonlocalHamiltonian} considered in the main text (see Fig.~\ref{fig:LanczosCoefficients}). We refer to this simplification of the Lanczos coefficients as the ramp-plateau model.

First, we will work in the thermodynamic limit and compute the location of the Krylov winding peak $\mu_K(t)$ and relate the peak width to the Lanczos growth rate $\alpha$. Then we will work with finite $N$ and compute the Krylov wavefunction numerically. The finite size data for $\mu_K(t)$ exhibits numerical scaling as a function of $N$ which we explain with a simple heuristic picture.

\subsection{Thermodynamic limit}

\begin{figure}[t!]
\centering
  \includegraphics[width=0.99\columnwidth]
 {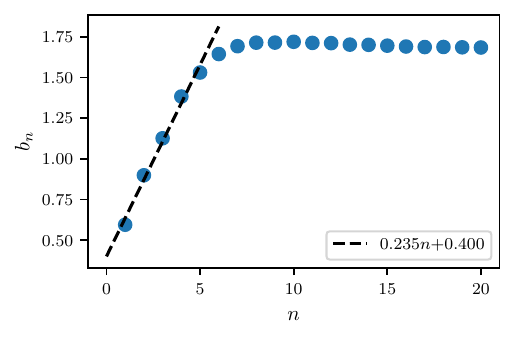}
	\caption{Disorder-averaged Lanczos coefficients of the nonlocal spin Hamiltonian~\eqref{eq:NonlocalHamiltonian} in the main text with $N=8, \beta=1$, and $\mO=S_{1}^{x}$. The Lanczos coefficients are approximately described by a linear ramp at small $n$; a linear numerical fit is shown (dashed line). The coefficients quickly saturate to a plateau which is approximately constant for $n\ll 4^{N}$. }
\label{fig:LanczosCoefficients}
\end{figure}

First, let us consider the limit $N\to\infty$ so that the Krylov wavefunction is described by the exact solution
\begin{equation}\label{eq:SupExactLinearWavefunction}
\varphi_{n}(t_\beta)=\frac{\tanh\left[\LanczosSlope t_\beta\right]^{n}}{\cosh\left[\LanczosSlope t_\beta\right]}
\end{equation}
where we have defined $t_\beta = t+i\beta/4$. With this solution the Fourier transform $C_{K}(t_\beta,\mu)$ is exactly computable,
\begin{equation}
\begin{aligned}
C_{K}(t_\beta,\mu) &= \sum_{n=0}^{\infty}\varphi_{n}^2(t_\beta)e^{in\mu}
\\
&=\frac{1}{\left(1-e^{i\mu}\tanh^2(\alpha t_\beta)\right)\cosh^2(\alpha t_\beta)}
\end{aligned}
\end{equation}
This function has a pole defined by the condition $e^{i\mu} = \coth^2(\alpha t_\beta)$. In general, there is no (real) choice of $\mu$ which satisfies this constraint; even so, it remains the case that $C_K$ is peaked around the frequency
\begin{equation}\label{eq:muKThermoLimitSM}
\mu_K(t_\beta) = - 2 \alpha_K(t) = -2\text{Arg}\left[\tanh(\alpha t_\beta)\right]
\end{equation}
This result was already used in Fig.~\ref{fig:Krylov_FTs}(c) of the main text with a numerical value for $\alpha$ obtained from the fit shown in Fig.~\ref{fig:LanczosCoefficients}.

The shape of the Krylov winding peak about $\mu=\mu_K(t_\beta)$ can also be determined in a straightforward manner. Setting $\mu=\mu_{K}(t)+\delta\mu$ and expanding $|C_K|^2$ to quadratic order,
\begin{equation}
|\cosh^2(\alpha t_\beta)|^2|C_K|^2=\frac{(1-|\tanh(\alpha t_\beta)^2|)^2}{1+\delta\mu^2\frac{|\tanh(\alpha t_\beta)^2|}{\left(1-|\tanh^2(\alpha t_\beta)|\right)^2}}
\end{equation}
This describes a Lorentzian of width
\begin{equation}
\begin{aligned}
\Delta\mu^2 &= \frac{\left(1-|\tanh^2(\alpha t_\beta)|\right)^2}{|\tanh(\alpha t_\beta)^2|}
\\
&=-2+\left|\coth^2(\alpha t_\beta)\right|+\left|\tanh^2(\alpha t_\beta)\right|
\end{aligned}
\end{equation}
which decays exponentially in time.

As a final comment on the thermodynamic limit, we recall the known thermal bound on the growth of Lanczos coefficients $\alpha\leq \pi/\beta$~\cite{hyp}. When this bound is saturated, the Lorentzian width vanishes, $\Delta\mu=0$, and the Krylov distribution remains perfectly peaked for all $t$.

\subsection{Finite $N$}

\begin{figure}[t!]
\centering
  \includegraphics[width=0.99\columnwidth]
 {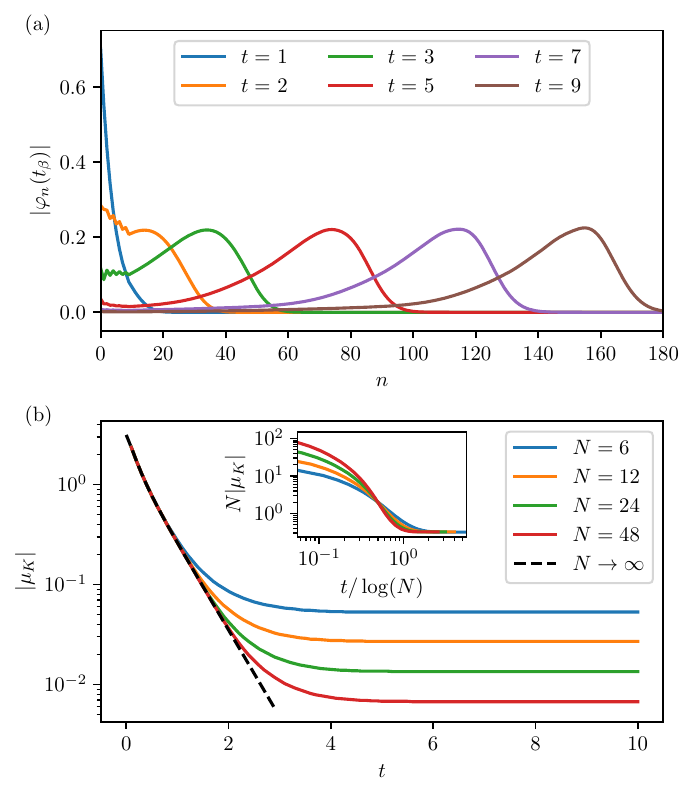}
	\caption{Numerically-obtained results for the ramp-plateau model with $\beta=\alpha=1$. (a) Krylov wavefunctions for $N=10$. For $t \gg \alpha^{-1} \log N$, the Krylov wavefunction approximately retains its shape and propagates ballistically. (b) The (absolute value of the) location of the peak in $C_K(\mu,t)$, as a function of $t$ and $N$. For $t \gg \alpha^{-1} \log N$, $\mu_K$ approaches a non-zero value due to finite size effects. The $N\to\infty$ result is given by~\eqref{eq:muKThermoLimit}. Inset: The rescaled winding peak, $N|\mu_K|$, vs. $t/\log(N)$. This rescaling collapses the scrambling time and late time plateaus.}
\label{fig:KrylovWavefunctions}
\end{figure}

Here we consider the ramp-plateau model~\eqref{eq:RampPlateauDef} for finite $N$. Prior to the time $\tscramb =\alpha^{-1}\log(N)$, the Krylov wavefunction is effectively confined to the linear ramp and accurately described by the exact solution~\eqref{eq:SupExactLinearWavefunction}. The support of the Krylov wavefunction expands exponentially in time with a rate determined by $\alpha$, which is well-known from prior studies of Krylov complexity in chaotic systems~\cite{hyp}.

Beyond $\tscramb$, the Krylov wavefunction enters the plateau and exhibits qualitatively distinct behavior. Through exact diagonalization of the Liouvillian we have found that the wavefunction in this regime develops a well-defined shape and propagates ballistically (see Fig.~\ref{fig:KrylovWavefunctions} (a)).

The time $\tscramb$ is also an important timescale for the behavior of the Krylov peak $\mu_K$ (see Fig.~\ref{fig:KrylovWavefunctions} (b)). Beyond $\tscramb$, the location of the peak in $C_K(\mu)$ saturates to a plateau value which scales as $1/N$. This scaling follows immediately from the exact result for the thermodynamic limit~\eqref{eq:muKThermoLimit} and the assumption that the time $t_{\ast}=\alpha^{-1}\log(N)$ signals the onset of finite size effects. The scaling collapse shown in the inset of Fig.~\ref{fig:KrylovWavefunctions} demonstrates that this is indeed the case.

\end{document}